\documentclass[aps,prb,10pt,a4paper,citeautoscript,superscriptaddress]{revtex4-2}

\usepackage{amssymb}
\usepackage{tikz}
\usepackage{natbib}
\usepackage{mathtools}
\usepackage{hyperref}
\usepackage{color}
\hypersetup{colorlinks=true, allcolors=blue}
\usepackage[T1]{fontenc}
\usepackage[top=2cm, bottom=2cm, right=2cm, left=2cm]{geometry}

\DeclareGraphicsExtensions{.pdf,.png,.gif,.jpg}

\newcommand{\Bayreuth}{Theoretische Physik III, Universit{\"a}t Bayreuth, 95440 Bayreuth, Germany}
\newcommand{\Dortmund}{Condensed Matter Theory, Department of Physics, TU Dortmund, 44221 Dortmund, Germany}
\newcommand{\Vienna}{University of Vienna, Faculty of Physics, Vienna Center for Quantum Science and Technology (VCQ), Vienna, Austria}

\begin{document}

\title{Photon Number Coherence in Quantum Dot-Cavity Systems can be Enhanced by Phonons}
\author{Paul C. A. Hagen}
\affiliation{\Bayreuth}
\author{Mathieu Bozzio}
\affiliation{\Vienna}´
\author{Moritz Cygorek}
\affiliation{\Dortmund}
\author{Doris E. Reiter}
\affiliation{\Dortmund}
\author{Vollrath M. Axt}
\affiliation{\Bayreuth}

\begin{abstract}
\noindent
    Semiconductor quantum dots are a versatile source of single photons with tunable properties to be used in quantum-cryptographic applications. A crucial figure of merit of the emitted photons is photon number coherence (PNC), which impacts the security of many quantum communication protocols. In the process of single-photon generation, the quantum dot as a solid-state object is subject to an interaction with phonons, which can therefore indirectly affect the PNC. In this paper, we elaborate on the origin of PNC in  optically excited quantum dots and how it is affected by phonons. In contrast to the expectation that phonons always deteriorate coherence, PNC can be increased in a quantum dot-cavity system due to the electron-phonon interaction. 
\end{abstract}

\maketitle

\section{Introduction}\label{sec:introduction}
Quantum dots (QDs) have been shown to be excellent single-photon sources for many applications and have steadily been improved regarding certain figures of merit, such as purity,\cite{Kiraz2004, Cosacchi2019, Bozzio2022, Michler2000, Giesz2016, Wei2014} indistinguishability \cite{Wei2014, Sbresny2022, Vajner2024} and brightness.\cite{Cosacchi2019, Bozzio2022, Michler2000, Wei2014} Another interesting characteristic of the emitted photons is the coherence between different photon-number states, known as photon number coherence (PNC). So far, PNC has not been much discussed in the quantum-dot community until it was discovered that PNC is of utmost importance for the performance of many quantum-cryptographic protocols \cite{Karli2024, Basset2021, Vajner2022}. Some protocols, such as the famous BB84 key distribution protocol,\cite{Bennett2014} achieve the highest security when PNC is as low as possible,\cite{LoPreskill2007, DusekJahmaLuetkenhaus2000} while other protocols require the presence of finite PNC,\cite{Bozzio2022,Karli2024} such as twin-field quantum key distribution with single photons.\cite{Lucaramini2018} Thus, the optimal value of PNC depends on the specific cryptographic protocol \cite{Bozzio2022} and how much deviation the protocol can tolerate.\cite{Karli2024, Basset2021, Vajner2022} PNC has so far mostly been investigated using Poisson sources.\cite{DusekJahmaLuetkenhaus2000,  LoPreskill2007} In most cases, achieving low PNC is the target. Thus, for Poisson sources usually a separate phase-scrambling step is performed,\cite{Cao2015, Zhang2021} which additionally decreases the secure key rate and can enable unambiguous state discrimination attacks, if it is performed imperfectly.\cite{Kobayashi2014, DusekJahmaLuetkenhaus2000} Additionally, Poisson sources must be used at very low brightness to work in the few-photon regime,\cite{Brassard2000} while QDs provide photons on-demand \cite{schweickert2018demand} promising a significantly higher communication rate.\cite{Zhang2024}\\

A drawback of QDs as photon emitters is often their interaction with the solid-state environment, in particular with phonons. Phonons are known to induce dephasing in QDs, i.e, they typically destroy coherence,\cite{Reiter2014, Skinner1986, Wang1998, Takagahara1999, Krummheuer2002, Krummheuer2005, Kaer2010} lead to reduced excited state preparation \cite{RamsayEtAl2010b,luker2019review} and deteriorate photonic properties.\cite{Cosacchi2021} For PNC, the influence of phonons is less obvious, as phonons do not act on the photons directly, but only affect the QD coherence.\\

In this paper, we study the influence of phonons on the PNC in a laser driven QD-cavity system. After establishing the electronic coherence as the origin of PNC, we discuss losses and quantify the phonon influence up to temperatures of $\sim 100$~K. The typical expectation is that phonons will destroy coherence, as known from the electronic coherence. In contrast to this expectation, we find that for certain parameters, phonons can drastically change the expected behaviour of PNC and even increase it compared to the phonon-free case.

\section{The QD-cavity system}\label{sec:system}
We consider an InGaAs/GaAs QD which we treat as a two-level system and is embedded in a mircocavity and is optically driven by a laser pulse. The uncoupled energies of the QD and the cavity are given by 
\begin{equation}
    \hat{H}^{\text{QD}+\text{cav}} = \hbar \Delta\omega_{XL}\vert X\rangle\langle X\vert + \hbar \Delta\omega_{CL}\hat{a}^\dagger \hat{a}, 
\end{equation}
in the rotating frame of the laser. The state $\vert X\rangle$ describes the exciton and has the energy $\omega_X$. $\vert G\rangle$ denotes the QD ground state, the energy of which is set to zero, and the energy $\Delta\omega_{XL} = \omega_X - \omega_L$ is the energy difference between the exciton and the laser. Similarly, $\Delta\omega_{CL} = \omega_C - \omega_L$ describes the difference between cavity mode and laser energy. $\hat{a}$ ($\hat{a}^\dagger$) is the photonic annihilation (creation) operator of the cavity, which is coupled to the QD with the coupling strength \mbox{$\hbar g = 0.05~\text{meV}$},\cite{KistnerEtAl2010, Hopfmann2015, YoshieEtAl2004} by \cite{Nazir2016}
\begin{equation} \label{eq:QD-cav-coupl.Hamilton}
    \hat{H}^{\text{int}} = \hbar g\left(\hat{a}^\dagger \vert G\rangle\langle X\vert + \hat{a} \vert X\rangle\langle G\vert\right).
\end{equation}
We label the QD-cavity states $\vert S\,n\rangle$, where $n\in \mathbb{N}_0$ refers to a photonic state with $n$ photons and $S\in\{G,X\}$ signifies the QD state.  
The cavity-exciton detuning is $\Delta\omega_{CX} = \omega_C - \omega_X$. A classical laser pulse is used to excite the QD. It is described by the time-dependent pulse-envelope function $f(t)$ and adds the term \cite{Nazir2016}
\begin{equation}
    \hat{H}^{\text{laser}} =  -\hbar \frac{f(t)}{2}\left(\vert G\rangle\langle X\vert + \vert X\rangle\langle G\vert\right)
\end{equation}
to the Hamiltonian. Gaussian pulses of the form
\begin{equation}
    f(t) = \frac{\Theta}{\sqrt{2\pi}\,\sigma}\text{e}^{-\frac{t^2}{2\sigma^2}},
\end{equation}
with a full width at half maximum $\text{FWHM} = 2\sqrt{2\text{ln}\,2}\,\sigma = 3~\text{ps}$ of the laser amplitude, and a pulse area $\Theta$, are used. Since it is embedded into the solid state environment, the quantum dot interacts with phonons, which provide an important dephasing mechanism. This coupling can generally occur due to different coupling mechanisms, in particular Fröhlich coupling to longitudinal optical (LO) phonons, piezoelectric coupling to all acoustic phonons as well as the deformation potential coupling to longitudinal acoustic (LA) phonons.\cite{luker2019review} Because the electron and hole distributions in neutral QDs are usually similar, the piezo-electric and Fröhlich couplings are very small and do not produce significant dephasing in GaAs QDs. \cite{Krummheuer2002, Krummheuer2005} For these reasons, and since LO phonons are assumed to be off-resonant to the transitions to higher electronic QD states, which would otherwise form a polaron in the QD,\cite{Hameau1999} we consider only the deformation potential coupling to LA phonons. The LA phononic environment, where $\hat{b}_j$ ($\hat{b}^\dagger_j$) annihilates (creates) a phonon of mode $j$ with energy $\omega_j$, is described by the Hamiltonian \cite{Besombes2001,Machnikowski2004}
\begin{equation} \label{eq:phononHamiltonian}
    \hat{H}^{\text{phon}} = \hbar\sum_{j}\omega_j\hat{b}^\dagger_j\hat{b}_j + \hbar \sum_{j}\left( \gamma_j^X\hat{b}_j^\dagger + \gamma_j^{X^*}\hat{b}_j\right)\vert X\rangle\langle X\vert,
\end{equation}
coupling each mode to the exciton with the coupling constant $\gamma_j^X$.\cite{Krummheuer2002,Krummheuer2005} It should be noted that Equation~(\ref{eq:phononHamiltonian}) is the standard microscopic basis to model phonon-induced processes in QDs, in particular pure dephasing.\cite{Krummheuer2002} To calculate the coupling constants, we use a simple model in which it is assumed that the electron and the hole are exposed to a harmonic potential. Then, the carrier wavefunctions are given by Gaussians, which have a characteristic width of $a_{e/h}$ for electron and hole, respectively. The resulting QD-phonon coupling constants are given by \cite{Krummheuer2002, luker2019review}
\begin{equation}
\gamma_j^X = \frac{|\boldsymbol{q}_j|}{\sqrt{2\hbar\rho\omega_j}}\left(D_e\,\text{e}^{-\boldsymbol{q}_j^2a_e^2/4} - D_h\,\text{e}^{-\boldsymbol{q}_j^2a_h^2/4}\right),
\end{equation}
where $D_{e/h}$ are the deformation coupling constants, $\rho$ is the mass density of the material and $\boldsymbol{q}_j$ is the wave-vector of the $j$th phonon mode, which has a frequency of $\omega_j$. This model has given excellent agreement with previous experiments with high sensitivity to the QD-phonon coupling.\cite{Quilter2015} We chose $a_e = 3~\text{nm}$, which corresponds to a FWHM of the electron density of 7~nm. Here, we assume the same harmonic potential for holes, resulting in $a_h = \frac{m_h}{m_e}\,a_e \approx 1.15\,a_e$. It has been found that it is also possible to describe non-spherical QDs using this formula when $a_{e/h}$ are appropriately chosen.\cite{luker2017}
All other phonon parameters of the InGaAs/GaAs QD were chosen as in Ref.~\cite{Barth2016}. Thus the total Hamiltonian is given by
\begin{equation} \label{eq:fullHamiltonian}
    \hat{H} = \hat{H}^{\text{QD}+\text{cav}} + \hat{H}^{\text{int}} + \hat{H}^{\text{laser}} + \hat{H}^{\text{phon}}.
\end{equation}

\begin{figure}[h]
    \centering
    \includegraphics[scale=0.75]{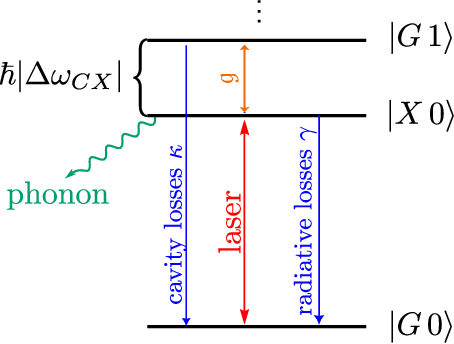}
  \caption{Energy diagram of the most relevant states of the QD-cavity system $|G0\rangle$, $|G1\rangle$ and $|X0\rangle$. The latter two states are coupled via the cavity coupling $g$, while $|G0\rangle$, $|X0\rangle$ are coupled via the laser. Loss channels are indicated by blue arrows. }
  \label{fig:energyLevels}
\end{figure}

Radiative losses of the QD and cavity losses are described by Lindblad-superoperators \cite{Lindblad1976, Carmichael1993} affecting the density operator $\hat{\rho}$
\begin{equation}
    \mathcal{L}_{\hat{O}, \delta}[\hat{\rho}] = \delta\left( \hat{O}\hat{\rho}\,\hat{O}^\dagger - \frac{1}{2}\left[\hat{\rho}, \hat{O}^\dagger \hat{O}\right]_+\right),
\end{equation}
where $\hat{O}$ is an operator, $\delta$ is the decay rate and $[\,\cdot\,,\,\cdot\,]_+$ the anti-commutator. The rates used for the radiative and cavity losses are $\gamma = 0.001~\text{ps}^{-1}$ and $\kappa  = 0.577~\text{ps}^{-1}$, respectively.
The dynamics are given by the Liouville-von Neumann equation
\begin{equation} \label{eq:liouvilleVonNeumann}
    \frac{\text{d}}{\text{d}t}\hat{\rho} = -\frac{\text{i}}{\hbar}\left[\hat{H},\hat{\rho}\right] + \mathcal{L}_{\vert G\rangle\langle X\vert,\gamma}\left[\hat{\rho}\right] + \mathcal{L}_{\hat{a},\kappa}\left[\hat{\rho}\right],
\end{equation}
where $[\,\cdot\,,\,\cdot\,]$ denotes the commutator. Solving the phonon system and its influence on the photon properties requires state-of-the-art methods, as otherwise qualitative and quantitative errors can occur.\cite{Cosacchi2021}\\
In order to obtain specific predictions starting with a Hamiltonian
of the type Equation~(\ref{eq:fullHamiltonian}) numerous methods are in use, like, e.g., non-equilibrium Green's function \cite{Hornecker2017},
the correlation expansion of density matrices \cite{Foerster2003,ForesterEtAl2003,Krugel2006,Reiter2017},
time-convolutionless projection operators \cite{Breuer2002} or the polaron master equation
\cite{McCutcheon2010,Nazir2016} to name but a few. Most of these methods introduce approximations to the model to make the numerics feasible. For example, the correlation expansion accounts only incompletely for multi-phonon processes which leads, e.g., to missing structures in corresponding spectra \cite{Krummheuer2002}. The correlation expansion also has limits when strong phonon coupling or long simulation times are considered \cite{Vagov2011_2}. The polaron master equation involves a Markov approximation in the polaron transformed frame which makes this approach unreliable at strong time-dependent driving. As a result, the standard polaron master equation does not correctly capture, e.g. the reappearance of Rabi rotations \cite{Nazir2016} as predicted in Ref.~\cite{Vagov2007} that has recently been measured \cite{Hanschke2024}.\\

Indeed, it is  challenging to account for the wealth of phenomena
induced by the QD-phonon coupling that go far
beyond simple pure-dephasing obtained by weak-coupling master
equations. This includes phonon-assisted transitions and thermalization,
polaron formation, which dresses and renormalizes driving and coupling
terms in the QD Hamiltonian, and non-Markovian quantum dynamics.
Moreover, also the effective dephasing rate strongly depends on details
of the driving, which is particularly challenging to work with in the
case of pulsed driving.\\

For simple systems, such as a two-level system coupled to an oscillator bath, the pioneering work of Makri and Makarov \cite{Makri1995_1,Makri1995_2} has shown that a numerical treatment without approximation to the model is possible by representing the time-evolution operator of the system formally exact as a path integral and performing the necessary sum over the paths using an iterative scheme. For the model considered in the present paper, however, the original algorithm would require the iteration of estimated $\sim 10^{11}$ complex valued numbers. Thus, we use for our numerics a modified algorithm \cite{Cygorek2017} that reduces in our case the number of elements that need to be iterated by about 5 orders of magnitude. Our modified path-integral approach can be formulated in Liouville-space \cite{Barth2016}, as needed when Markovian rates e.g. for cavity losses and radiative decay should be accounted for, as we do in Equation~(\ref{eq:liouvilleVonNeumann}). It is also suitable for calculating multi-time correlation functions \cite{Cosacchi2018}, which is necessary for the calculation of the purity and indistinguishability. Our implementation of path-integrals for the simulation of multi-time functions avoids the use of the quantum regression theorem (QRT). The QRT involves a Markov approximation even when the memory for the single time propagations is fully accounted for, which can introduce sizable errors.\cite{Cosacchi2021, Cosacchi2021Erratum}
We are able to obtain converged results that invoke no approximation to the model such that
only well controlled discretization errors occur.\\

The QD density matrix is calculated by tracing out the cavity's and phononic degrees of freedom
\begin{align} \label{eq:photonicRDM}
    \rho_{\mu\nu} \coloneqq \langle \mu| \text{Tr}_{\text{cav}}\,\text{Tr}_{\text{phon}}(\hat{\rho}) |\nu\rangle  && \mu,\nu\in\{G,X\},
\end{align}
and similarly, the reduced photonic density matrix is obtained by
\begin{align}
    \rho_{ij} \coloneqq\langle i| \text{Tr}_{\text{QD}}\,\text{Tr}_{\text{phon}} (\hat{\rho}) |j\rangle && i,j \in \mathbb{N}_0,
\end{align}
where $|j\rangle$ denotes a state with $j$ photons in the cavity.\\

We focus on the experimentally relevant regime for single-photon emission, where the total excitation number is limited. The most relevant states are the energetically lowest three levels $\vert G0\rangle$, $\vert G1\rangle$ and $\vert X0\rangle$ which are displayed in \textbf{Figure~\ref{fig:energyLevels}}. In our numerical simulations, states with up to $n = 2$ photons are accounted for. Because under our excitation conditions, occupations of  cavity photon states with more than one photon are orders of magnitude smaller than the single-photon state, we henceforth focus our discussion on the photon number coherence between the vacuum and single-photon states $\rho_{01}$.

\section{The origin of photonic coherence}\label{sec:origin}
The laser excitation leads to coherence between the ground state and the excited state, which we will refer to as $\rho_{GX}$. Using the Heisenberg equation of motion of the photonic coherence $\rho_{01}$, one finds
\begin{equation} \label{eq:rho01ODE}
    \frac{\text{d}}{\text{d}t}\rho_{01}(t) = \left(-\text{i}\Delta\omega_{CL}-\frac{\kappa}{2}\right)\rho_{01}(t) - \text{i}g\rho_{GX}(t).
\end{equation}
The equation demonstrates that $\rho_{01}$ represents a driven oscillator with a free oscillation frequency $\Delta\omega_{CL}$, that is damped via the rate $\kappa/2$. Most importantly, the coherence is driven by the QD coherence $\rho_{GX}$, hence the electronic coherence $\rho_{GX}$ is the source of photonic coherence $\rho_{01}$.\\
The solution of Equation~(\ref{eq:rho01ODE}) is formally given by
\begin{equation} \label{eq:LorentzFilter}
    \rho_{01}(t) = -\text{i}g \int_{-\infty}^t \rho_{GX}(t')\,\text{e}^{\left(-\text{i}\Delta\omega_{CL}-\frac{\kappa}{2}\right)(t-t')}\text{d}t',
\end{equation}
linking the photon coherence to all past QD coherences by a memory function that constitutes a Lorentzian filter with a peak transmission at $\hbar\Delta\omega_{CL}\approx 0~\text{meV}$ and a band width determined by $\kappa/2$. Note that Equation~(\ref{eq:LorentzFilter}) remains exact when phonons couple to the QD because the QD-phonon interaction commutes with all cavity operators. Yet, phonons can still affect the PNC via the QD coherences $\rho_{GX}$.\\

\begin{figure}
    \centering
    \includegraphics[scale = 1]{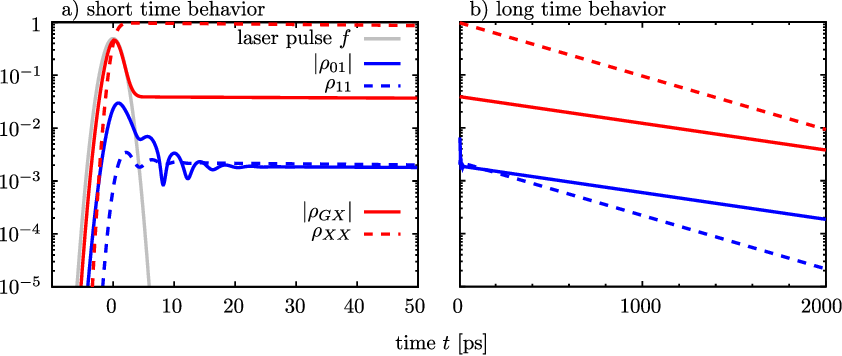}
  \caption{a) Short-time and b) long-time dynamics of the system excited by a $\pi$-pulse. Displayed is the QD population $\rho_{XX}$ and electronic coherence $|\rho_{GX}|$, single-photon state population $\rho_{11}$ and the photonic coherence $|\rho_{01}|$ for a QD-cavity detuning of $\hbar\Delta\omega_{CX}=1~\text{meV}$. These calculations were done without phonons, but the coherence behavior does not change qualitatively when phonons are included.}
  \label{fig:temporalBehaviour}
\end{figure}

The time dependence of the absolute values of the coherences $|\rho_{01}|$ and $|\rho_{GX}|$ alongside the populations $\rho_{11}$ and $\rho_{XX}$ are shown in \textbf{Figure~\ref{fig:temporalBehaviour}} for an excitation with a $\pi$-pulse and a QD-cavity detuning of 1~meV. The pulse $f(t)$ is included for reference. The calculations were done without phonons. When a finite temperature and phonons are included, the qualitative behavior does not change. Three different dynamical regimes can be identified in the time evolution: The dynamics during the pulse, the short-time dynamics after the pulse and the long-time decay.\\

During the pulse, the system inverts, i.e., the population $\rho_{XX}$ goes to $\sim 1$, while the electronic coherence $|\rho_{GX}|$ reaches its maximum and then decreases again. During this maximum of the QD coherence, according to Equation~(\ref{eq:rho01ODE}) its emission into the cavity is large, such that $\rho_{01}$ increases quickly as well, producing the largest photonic coherence while the pulse is still active.\\

Just after the pulse, the photonic coherence drops off while displaying oscillations. These oscillations occur due to the detuning between QD and cavity, which is identical to $\Delta\omega_{CL}$ in Equation~(\ref{eq:rho01ODE}), because resonant pulses are used for the excitation. Because the stated $g$ and $\kappa$ values determine the system to be in the weak-coupling regime, these oscillations are only observed when adding a finite QD-cavity detuning $\hbar\Delta\omega_{CX}$. Additionally, a drop-off occurs on top of the oscillation.\\

After roughly \mbox{30 ps}, the system reaches its long-time regime, where the decay becomes monoexponential with a single decay rate $\lambda_+$.\\

The dynamics of the QD and photonic coherences after the pulse can be described by a simplified approximate model, which is presented in Appendix A. From this model we extract the short-time and long-time decay by a linear combination of two exponential functions
\begin{equation}\label{eq:analyticApprox}
    \rho_{01}(t) = A_+\,\text{e}^{\lambda_+t} + A_-\,\text{e}^{\lambda_-t}.
\end{equation}
with $\lambda_-$ standing for the initial faster decay and $\lambda_+$ for the long-time slower decay. In the phonon-free model, simple analytical expressions can be found for both decay rates and the amplitudes $A_{\pm}$. When phonons are included, Equation~(\ref{eq:analyticApprox}) still holds up to $\sim 70~\text{K}$, but $A_{\pm}$ and $\lambda_\pm$ must be determined numerically. The dependence of the decay rates $\lambda_\pm$ on the QD-cavity detuning and on temperature is discussed in Appendix B.

\section{Time-integrated PNC}
In most quantum-cryptographic studies, the details of the dynamics are not of interest.\cite{Bozzio2022, LoPreskill2007, Karli2024} Instead, a single figure of merit is desirable. A well defined quantity is the time integral over the absolute value of the reduced photonic density matrix
\begin{equation} \label{eq:rhoTildeDefinition}
    \tilde{\rho}_{ij} := \int_{-\infty}^\infty |\rho_{ij}|\,\text{d}t,
\end{equation}
with $i,j$ denoting the photon state. For $i\neq j$, this quantity is proportional to the PNC of the outside mode resonant with the QD-cavity system, once the photon has been fully emitted. Note that due to the integration, this measure of PNC does not distinguish between the two limiting cases, where the coherence is low at any given time, or is initially large and then decreases quickly.\\

We can use the fact that the integral in the calculation of the PNC from Equation~(\ref{eq:rhoTildeDefinition}) can be broken up into two components over the intervals $(-\infty,t_0)$ and $(t_0,\infty)$, where $t_0$ denotes the end of the laser pulse. The integral over the latter can then approximately be calculated by
\begin{equation} \label{eq:PNCApproxFRhoGX}
    \tilde{\rho}_{01}^{(t_0,\infty)} \approx F(\Delta\omega_{CX}, g)\,|\rho_{GX}(t_0)|,
\end{equation}
where $F$ is some filter function derived out of the filtering in Equation~(\ref{eq:LorentzFilter}) in combination with Equation~(\ref{eq:analyticApprox}). The derivation of this formula and an explicit analytical expression for $F$ can be found in Appendix A. Importantly, the filter function $F$ can not depend on any properties of the pulse. Instead, any dependence on, for example the pulse area, lies exclusively in the initial electronic coherence $|\rho_{GX}(t_0)|$. To understand the behavior of $\tilde{\rho}_{01}$, for which the integration in Equation~(\ref{eq:rhoTildeDefinition}) takes place over all times, it can be useful to consider only $\tilde{\rho}_{01}^{(t_0,\infty)}$, which containes the integration over most relevant times. Now, we are able to investigate the influence that changes in parameters, like the QD-cavity coupling or the pulse area, have by determining their influence on the filter function $F$ or the $|\rho_{GX}(t_0)|$. This simplifies the analysis greatly.\\

\begin{figure}
    \centering
    \includegraphics[scale = 1]{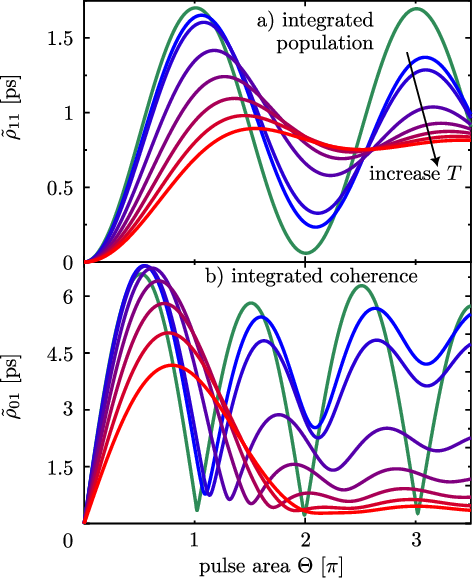}
  \caption{Time-integrated absolute values of a) the one photon population and b) the coherence between 0 and 1 photons, displayed in dependence on the laser pulse area $\Theta$ for different temperatures $T$. The chosen temperatures are \mbox{4.2 K} as well as \mbox{10 K} to \mbox{110 K} in \mbox{20 K} steps from blue to red. The phonon-free case is displayed in green. All calculations were done with QD-cavity detuning \mbox{$\hbar\Delta\omega_{CX} = 0~\text{meV}$}.}
  \label{fig:pulseAreaDependence}
\end{figure}

\subsection{Rabi rotations}\label{sec:rabiRotations}

\textbf{Figure~\ref{fig:pulseAreaDependence}} shows the integrated one-photon population $\tilde{\rho}_{11}$ along side the PNC $\tilde{\rho}_{01}$ as a function of pulse area $\Theta$ for different temperatures starting from $4.2$~K and then increasing from $10$~K in steps of $20$~K to $110$~K. For temperatures above$~80$~K, optical phonons are expected to play a prominent role,\cite{Carmele2010} restricting the validity of our model to temperatures $T\lesssim 100$~K. The phonon-free case is displayed in green for reference.\\

In the two-level system, without a cavity or any rate decay, varying $\Theta$ produces Rabi rotations in the QD population, meaning the QD is fully excited (relaxed) when $\Theta$ is an odd (even) multiple of $\pi$. In turn, the coherence $\rho_{GX}$ is, with an absolute value of 0.5, at its largest for $\Theta = \left(n+\frac{1}{2}\right)\pi, n\in\mathbb{N}_0$, where the QD is half excited, and zero, whenever the QD is fully excited or fully relaxed. Considering this behavior of the electronic coherence and the proportionality between the electronic coherence and PNC from Equation~(\ref{eq:PNCApproxFRhoGX}), this makes $\Theta$ a very effective tuning knob for PNC. Some deviations exist, in large parts due to the influence of the cavity, but the phonon-free case and the low temperature cases in Figure~\ref{fig:pulseAreaDependence} show the expected behavior. The deviations caused by the cavity are described in a model in Appendix C and produce for example the alternating height of the PNC maxima.\\

Phonons have several effects on the photon number and the PNC: A major effect is that the interaction with phonons leads to renormalization of the pulse area, shifting the laser intensity at which the QD is maximally excited towards larger values. Equally important is the effect that phonons dampen the Rabi rotations,\cite{Reiter2014,RamsayEtAl2010b,Kruegel2005,Vagov2011} resulting in a lower amplitude of the oscillations.\\

These effects can be observed in Figure~\ref{fig:pulseAreaDependence}. Focusing on the integrated occupation of the single photon state $\tilde{\rho}_{11}$ we can see both effects of renormalization and damping: The extrema shift towards higher pulse areas and the amplitude is decreased. For increasing temperature, the effective electron-phonon coupling becomes stronger and, correspondingly, the renormalization and damping becomes stronger.\\ 

These mechanisms similarly affect the electronic coherence and because of its close link to the photonic coherence, they also affect the PNC, which is shown in Figure~\ref{fig:pulseAreaDependence}~ b).\\
First, we consider $\pi$-pulses, where the PNC reaches almost zero in the phonon-free case. We find that the PNC remains much larger when phonons are accounted for. This effect is due to the damping of the Rabi rotations. Since a full inversion is not reached, the electronic coherence can still have significant values. And because the electronic coherence is the source of the PNC, the latter also increases.\\
When we instead consider $\pi/2$-pulses, where without phonons PNC is maximal, we observe that even here an increase of PNC is possible. This surprising effect can be traced back to the renormalized coupling strengths,\cite{Kaer2010,Hopfmann2015,Glaessel2012,MildeKnorrHughes2008} that influence this integrated quantity. The reason for this is that $F$ in Equation~(\ref{eq:PNCApproxFRhoGX}) increases for decreasing coupling and the coupling decreases due to the renormalization effect of phonons. Physically this is due to the slow-down in the emission of the coherence out of the QD-cavity system. Note that this effect of course plays a role for all pulse-areas, since $F$ does not depend on the initial excitation. This effect therefore also influences the increased value of PNC for $\pi$-pulses.\\

At higher pulse areas, the PNC is then strongly damped and no more oscillations are visible, if the temperature is sufficiently large. Thus, depending on the pulse area, phonons can have coherence boosting or damping effects.\\

\begin{figure}
    \centering
    \includegraphics[scale = 1]{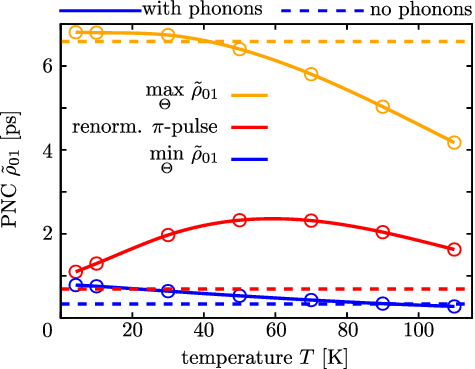}
  \caption{PNC for the pulse areas at which the QD occupation is maximized (red) and the PNC is maximized (orange) or minimized (blue), displayed for different temperatures. The respective phonon-free cases are displayed by the dashed lines. The calculations were done with $\hbar\Delta\omega_{CX} = 0~\text{meV}$. The orange and blue lines are the maximal and minimal values from Figure~\ref{fig:pulseAreaDependence}~b). The unit of the PNC is ps, because it is the integrated value.}
  \label{fig:minRabiRot}
\end{figure}

The impact of phonons on the first PNC extrema as a function a of temperature is summarized in \textbf{Figure~\ref{fig:minRabiRot}}, where the maximal and minimal values of PNC are displayed in orange and blue, respectively. At low temperatures, both increase when phonons are included. As the temperature increases, the phonon-induced pure dephasing takes over and decreases PNC below the phonon-free value.\\

The most interesting curve is the red one, which is for the renormalized $\pi$-pulse, i.e., when the QD excitation is maximized just after the pulse. Interestingly, while the occupation has its maximum, the PNC is not at its minimum. Again, this is an effect of the cavity, which is explained by the model in Appendix C. When we therefore consider the PNC at the renormalized $\pi$-pulse, we first find an increase of the PNC up to about 60~K, and then the PNC decreases.\\

\begin{figure*}
    \centering
    \includegraphics[scale = 1]{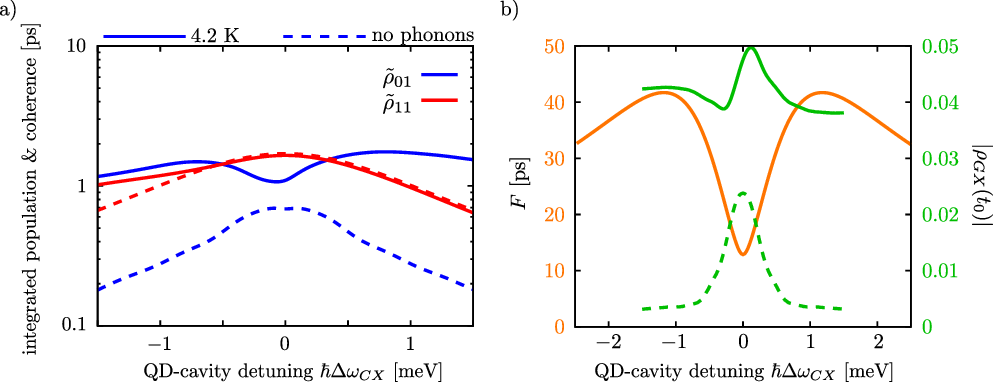}
  \caption{a) Time-integrated single-photon population $\tilde{\rho}_{11}$ and coherence $\tilde{\rho}_{01}$ as a function of the QD-cavity detuning $\hbar\Delta\omega_{CX}$ at $4.2~\text{K}$ (solid) and without phonons (dashed). Renormalized $\pi$-pulses were used for the calculation. b) Filter function (orange) as calculated in Appendix A, and the electronic coherences 
 at a time $t_0$, just after the pulse has ended (green). The solid (dashed) line denotes the case with (without) phonons.}
  \label{fig:finalPhotonDMs}
\end{figure*}

\subsection{QD-cavity detuning} \label{sec:finalDMDescription}
Another important tuning parameter for the PNC is the QD-cavity detuning. \textbf{Figure~\ref{fig:finalPhotonDMs}}~a) displays the resulting time-integrated one-photon population $\Tilde{\rho}_{11}$ and the 01-coherence $\tilde{\rho}_{01}$ for different QD-cavity detunings. $\tilde{\rho}_{01}$ increases by roughly one order of magnitude, when adding phonons and changes its shape. Whereas without phonons it has an almost imperceptible local minimum at \mbox{$\hbar\Delta\omega_{CX} = 0~\text{meV}$} and two maxima close to either side. The minimum is much more pronounced, when phonons are included.\\

To explain this phenomenon, we can again use the approximate model from Equation~(\ref{eq:PNCApproxFRhoGX}) and investigate the behavior of the filter function $F$ and the electronic coherence just after the pulse $|\rho_{GX}(t_0)|$, separately. Figure~\ref{fig:finalPhotonDMs}~b) displays the filter function (orange), which possesses a central minimum for the QD-cavity coupling displayed. In the case without phonons, the electronic coherence just after the pulse (dashed green), has a central peak which is narrow compared to the minimum of $F$. As a result, the product of the two, which approximates the PNC over the course of the emission, only has a very slight minimum in the middle, producing the phonon-free $\tilde{\rho}_{01}$ in subfigure~a). The QD coherence at time $t_0$ is generally much larger than in the phonon-free case \cite{RamsayEtAl2010b, Foerster2003, ForesterEtAl2003, Kruegel2005, Vagov2011} , which is again due to the damping described in Section~\ref{sec:rabiRotations}. As a consequence, the product of it and $F$, still possesses the pronounced central minimum.\\

As can be seen in Figure~\ref{fig:finalPhotonDMs}~a), in the case with phonons, $\tilde{\rho}_{01}$ is generally smaller for negative QD-cavity detunings, compared to positive ones, which is explained by phonon-induced transitions between QD and cavity. For negative detunings, phonons provide an incoherent and fast relaxation mechanism from the exciton to the cavity, which decreases the QD population and the coherence quicker. But while this leads to an increase of $\tilde{\rho}_{11}$, $\tilde{\rho}_{01}$ decreases due to the incoherent nature of the phonon-induced transition.\\

\begin{figure*}
    \centering
    \includegraphics[width = \linewidth]{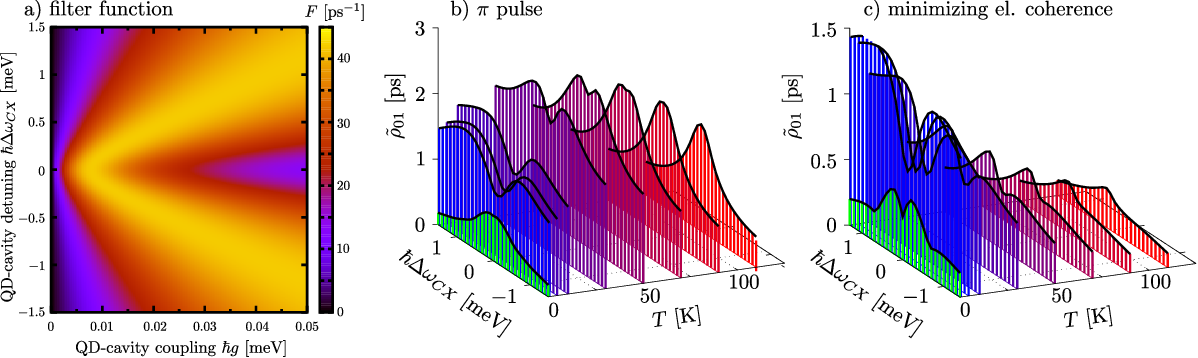}
  \caption{a) Dependence of the filter function $F$ of the detuning $\Delta\omega_{CX}$ and the QD-cavity coupling $g$ b)/c) Temporal integration of the photonic 01-number coherence, $\tilde{\rho}_{01}$, displayed for different temperatures $T$ and QD-cavity detunings $\hbar\Delta\omega_{CX}$. The chosen temperatures are \mbox{4.2 K} as well as \mbox{10 K} to \mbox{110 K} in \mbox{20 K} steps. The phonon-free case is displayed in green.\\
  The pulse area is chosen to b) maximize the QD population and to c) minimize the QD polarization after the pulse.}
  \label{fig:3DDifferentTempsPublication}
\end{figure*}

The temperature-dependence of the PNC can also be investigated using the filter function $F$. To do this, we need to first investigate the QD-cavity coupling dependence of $F$, which is displayed in \textbf{Figure~\ref{fig:3DDifferentTempsPublication}}~a). In the limit $\hbar g\to0~\text{meV}$, the filter function converges towards zero, since the cavity is hardly ever excited when such low coupling is present. For small finite coupling strengths, only a single central maximum of $F(\Delta\omega_{CX})$ exists. It turns into a central minimum, flanked by two maxima, if the coupling strength is increased above a certain threshold. It is well known that phononic effects of increasing temperatures decrease the QD-cavity coupling strength.\cite{Kaer2010,Hopfmann2015,Glaessel2012,MildeKnorrHughes2008} The effect of this can be seen in {Figure~\ref{fig:3DDifferentTempsPublication}}~b) and c), which display the PNC for different temperatures and QD-cavity detunings for a) pulse areas that maximize QD occupation just after the pulse ended and b) pulse areas that minimize electronic coherence just after the pulse ended.\\
These two plots show that the PNC for different temperatures is governed by the two effects already mentioned: The general behavior follows the one observed in Figure~\ref{fig:minRabiRot}. For example, the $\pi$-pulse produces an increase for small increasing temperatures, but then drops after roughly 55~K, whereas the pulse minimizing electronic coherence decreases the PNC monotonically for increasing temperatures, but also experiences phonon-induced damping as discussed in Section~\ref{sec:rabiRotations}. The other effect that temperature introduces is the aforementioned renormalizaton of the QD-cavity coupling, adding the central minimum for small temperatures, which disappeared independently from the pulse area, at roughly 55~K.\\

\begin{figure}
    \centering
    \includegraphics[width=0.5\linewidth]{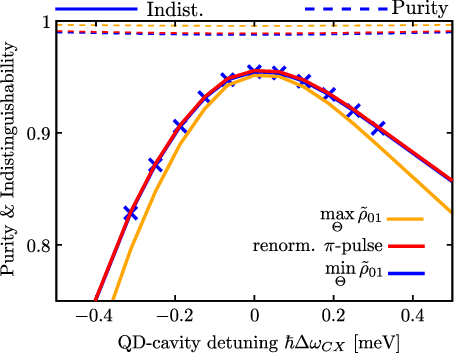}
    \caption{Purity (dashed) and indistinguishability (solid) for three different pulse areas, which correspond to Figure~\ref{fig:minRabiRot} for $4.2\text{ K}$. The pulse area that minimized PNC for $\Delta\omega_{CX}=0\text{ meV}$ is displayed in blue, the one corresponding to a $\pi$-pulse is shown in red, and the pulse area maximizing PNC is yellow. Blue crosses are added to make the data of the indistinguishability for minimal PNC discernible behind the red curve.}
    \label{fig:purityIndist}
\end{figure}

These results are encouraging despite the phonon-induced increase in PNC at low temperatures, when keeping other figures of merit of single-photon sources, such as purity, indistinguishability and brightness, in mind. In most cases, one is interested in having low PNC. While the minimally achievable PNC decreases for larger temperatures, the effect is not very pronounced in Figure~\ref{fig:minRabiRot} and one can reasonably use low temperatures without too large a penalty to PNC. In this case, very good values for purity, indistinguishability and brightness are possible \cite{Cosacchi2021, Somaschi2016} because the $\pi$-pulse area is approximately equal to the pulse area minimizing PNC for low temperatures. When instead, PNC should be maximized, one needs to use a pulse area similar to $\pi/2$. In this case, purity and indistinguishability still have high values, but the brightness suffers a significant penalty of about $1/2$.\cite{Somaschi2016} This reduction, however, must be accepted when electronic coherences are required e.g. for coherent control.\\
In order to quantify how important figures of merit perform for the parameters optimized for low or high PNC values, we have calculated purity and indistinguishability for the same parameters as in Figure~\ref{fig:minRabiRot} for $T=4.2\text{ K}$. Details how these quantities are calculated are given in Ref.~\cite{Cosacchi2021}. As seen in \textbf{Figure~\ref{fig:purityIndist}}, the purity is not significantly affected by changes in the QD-cavity detuning and stays near unity regardless whether PNC is optimized for high or low values. Its values are about $98\%$, which is typical for similar systems\cite{Cosacchi2021, Heinisch2024}. In particular, the results for $\pi$-pulses are almost the same as in the case minimizing PNC, which is likely due to the very similar pulse area needed for these cases. The pulse area maximizing PNC, which is roughly similar to the $\pi/2$ case, has higher purity because for lower driving strengths, the reexcitation probability is reduced. The indistinguishability reaches its highest values for QD-cavity detuning $\hbar\Delta\omega_{CX} = 0\text{ meV}$ and falls off for larger detunings in all considered cases. The indistinguishability values in Figure~\ref{fig:purityIndist} are in reasonable agreement with other results reported for similar systems.\cite{Cosacchi2021, Heinisch2024, Bauch2024} For cavity-based systems it has been reported in Ref.~\cite{Heinisch2024} that almost perfect indistinguishabilities and purities can be achieved by using the Swing-UP of quantum EmitteR (SUPER) \cite{Bracht2021} scheme for the preparation instead of Gaussian pulses.\\

These results demonstrate that for the parameters where the lowest PNC is reached, it is usually possible to simultaneously obtain high values for purity, indistinguishability and brightness. Also, when high PNC is the target, the achievable purity is generally rather high for low temperatures, while the indistinguishability has reasonable values that still depend on parameters like the QD-cavity detuning and details of the excitation. And naturally, for pulse areas near $\pi/2$ the brightness is significantly lower than for $\pi$-pulses.

\section{Conclusions}\label{sec:conclusion}
We have discussed the influence of phonons on the PNC of a single photon emitted from a QD. We started by establishing that the QD coherence is the source of PNC for a single photon source. Thus, all influences affecting electronic coherence, like losses or phonons, impact the PNC. \\

We have analyzed the PNC as function of pulse area, which for the QD occupation leads to the well-established Rabi rotations. In the phonon-free case, the control of the electronic system is passed on to the PNC. This changes significantly in the case with phonons. Dispite the well-known pure dephasing effects, phonons were found to sometimes increase PNC. For $\pi$-pulses the reason is the damping of Rabi rotations and for $\pi/2$-pulses it is mostly due to the renormalization of QD-cavity coupling, which slows the emission of the coherence.\\

Because of the renormalization of the QD-cavity coupling, different temperatures produce qualitatively different behavior of PNC for different QD-cavity detunings. Crucially, for low temperatures, the PNC increases for all pulse areas.\\

Our study shows that the influence of phonons on the PNC is highly non-trivial, as the simple assumption that the PNC exclusively decreases with phonons due to decoherence, does not hold true and other phononic effects must be taken into account. As PNC is crucial for many quantum-cryptography applications, it is essential to always consider the influence of the solid-state-environment, which might be detrimental, but also could turn out as helpful.\cite{Bozzio2022}\\

This implies that the strength of phonon coupling in applications benefiting from low PNC, such as standard/decoy BB84 QKD, unforgeable quantum tokens, quantum coin flipping and quantum bit commitment, should be carefully monitored to ensure practical security. On the other hand, phonons could provide better PNC in applications which require a phase reference such as twin-field quantum key distribution.\cite{Lucaramini2018}

\section*{Appendix A: Simple model of the relaxation}
It is helpful to develop an understanding of the phonon-free case to determine the impact of phonons on the resulting coherences. Because we find the most important photonic coherence to be the 01-coherence and because the coherence between $|G0\rangle$ and $|X0\rangle$ is  much larger than the one between $|G1\rangle$ and $|X1\rangle$, we can simplify the massive Hilbert space by looking only at the photonic coherence and the QD coherence, which are then mostly given by 
\begin{align}
	\rho_{01} &= \rho^{G0}_{G1} + \rho^{X0}_{X1}\\
	\rho_{GX} &\approx \rho^{G0}_{X0},
\end{align}
where the short notation
\begin{equation}
    \rho_{S_2n_2}^{S_1n_1} = \langle S_1n_1\vert \text{Tr}_{\text{phon}}(\hat{\rho})\vert S_2n_2\rangle
\end{equation}
was used. Through the Liouville-von-Neumann equation Equation~(\ref{eq:liouvilleVonNeumann}) these two coherences are coupled to each other. For times when the laser is not active and if the phonon-free case is considered, these equations read
\begin{equation} \label{eq:coherenceDGL}
    \frac{\text{d}}{\text{d}t}\left[\begin{array}{c}\rho_{GX}\\ \rho_{01} \end{array}\right] =
    \left[\begin{array}{cc}
        -\text{i}\Delta\omega_{XL}-\frac{\gamma}{2} & -\text{i} g \\
        -\text{i} g & -\text{i}\Delta\omega_{CL}-\frac{\kappa}{2}
    \end{array}\right]
    \left[\begin{array}{c} \rho_{GX}\\ \rho_{01} \end{array}\right].
\end{equation}
The behavior of the system can then be predicted by solving this ordinary differential equation (ODE) and depends on only two initial values. Let $t_0>0$ be the time at which the laser intensity after the maximum is sufficiently small to provide these initial values $\rho_{01}(t_0)$ and $\rho_{GX}(t_0)$, then the solution of Equation~(\ref{eq:coherenceDGL}) can be written in the form given in Equation~(\ref{eq:analyticApprox}) in the main text. From the solution of this ODE, we obtain the squared absolute photonic coherence
\begin{equation}\label{eq:meanPhotonicCoherenceAnalyticApproxAppendix}
    \begin{split}
        |\rho_{01}(t)|^2 &= \frac{1}{|\text{det}\,T|^2}\Big(A(\Delta\omega_{CX})\,\text{e}^{2\text{Re}\lambda_+(t-t_0)}\\ &+2\,\text{Re}\left[B(\Delta\omega_{CX})\,\text{e}^{(\lambda_+ + \lambda_-^*)(t-t_0)} \right]\\ &+C(\Delta\omega_{CX})\,\text{e}^{2\text{Re}\lambda_-(t-t_0)} \Big),
    \end{split}
\end{equation}

\begin{figure}
    \centering
    \includegraphics[scale = 1]{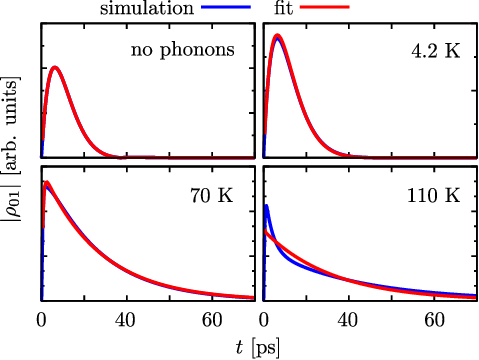}
  \caption{Comparison between full simulation and best fit of linear combination of two exponential functions, as in Equation~\ref{eq:analyticApprox} in the main text, for different temperatures. These calculations were done using an initial value problem with a fully excited QD.}
  \label{fig:fitExamples}
\end{figure}

with
\begin{align}
    A(\Delta\omega_{CX}) &= |\rho_{01}(t_0)-\rho_{GX}(t_0)T_+|^2|T_-|^2\label{eq:factorA}\\
    \begin{split}
    B(\Delta\omega_{CX}) &= \Big(-|\rho_{GX}(t_0)|^2T_-^*T_+ - |\rho_{01}(t_0)|^2\\
    &+ \rho_{GX}(t_0)\rho_{01}^*(t_0)T_+\\
    &+ \rho_{GX}^*(t_0)\rho_{01}(t_0)T_-^*\Big)\,T_-T_+^*
    \end{split}\\
    C(\Delta\omega_{CX})& = |\rho_{01}(t_0)-\rho_{GX}(t_0)T_-|^2|T_+|^2
\end{align}
and
\begin{align}
    Q &= \sqrt{\left(\text{i}\Delta\omega_{CX}+\frac{\kappa-\gamma}{2}\right)^2 - 4g^2}\\
    \text{det}\,T &= T_+-T_- = -\frac{\text{i}}{g}Q\\
    \lambda_{\pm} &= \frac{-\text{i}(\Delta\omega_{XL}+\Delta\omega_{CL})-\frac{\kappa+\gamma}{2}\pm Q}{2} \label{eq:ReLambda1}\\
    T_{\pm} &= \frac{\Delta\omega_{CX}}{2g}-\frac{\text{i}}{2g}\left( \frac{\kappa-\gamma}{2} \pm Q\right).
\end{align}
\textbf{Figure~\ref{fig:fitExamples}} shows that such a model is applicable even when phonons are active up to about 70~K. In that case, the decay rates must be fitted as they are affected by e.g. phonon renormalizations of the QD-cavity coupling, a polaron shift and a pure dephasing rate.

\begin{figure}
    \centering
    \includegraphics[scale = 1]{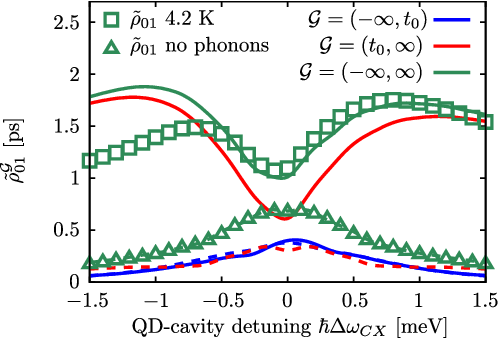}
  \caption{The resulting integrated PNCs according to Equation (\ref{eq:integratedCoherenceAfterPulse},\ref{eq:integratedCoherenceOverall}) for three different intervals $\mathcal{G}$: during the pulse $\mathcal{G} = (-\infty,t_0)$, after the pulse $\mathcal{G} = (t_0,\infty)$ and over the entire duration $\mathcal{G} = (-\infty, \infty)$. Additionally, the simulation results for $\tilde{\rho}_{01}^{(-\infty,\infty)}$ are displayed for $4.2~\text{K}$ (square) and without phonons (triangle).}
  \label{fig:analyticFactor}
\end{figure}

In case accurate values for $\tilde{\rho}_{01}$ are required, $\sqrt{|\rho_{01}(t)|^2}$ can be integrated numerically. However, since the integral is dominated by the slow exponential decay after the oscillations have ceased, we can approximate
\begin{equation}
    |\rho_{01}(t)|^2 \hspace{.5em} \stackrel{t-t_0 >\!\!> 2/\kappa}{\approx} \hspace{.5em} \frac{A(\Delta\omega_{CX})\,\text{e}^{2\text{Re}\lambda_+(t-t_0)}}{|\text{det}\,T|^2}.
\end{equation}
Eventually, integration with Equation~(\ref{eq:factorA}) and using $|\rho_{01}(t_0)-\rho_{GX}(t_0)T_+| \approx |\rho_{GX}(t_0)T_+|$, which applies because $|T_+|\sim 10$ and because the initial electronic coherence is larger than the photonic one, results in an approximate formula for the integrated coherence after the excitation. To shorten notation, we introduce
\begin{equation}
    \tilde{\rho}_{01}^\mathcal{G} := \int_{\mathcal{G}} |\rho_{01}(t)|\,\text{d}t
\end{equation}
for arbitrary integration intervals $\mathcal{G}$. Because of Equation~(\ref{eq:rhoTildeDefinition}) $\tilde{\rho}_{01}(t_0) = \tilde{\rho}_{01}^{(-\infty,\infty)}$ applies. The stated approximations result in
\begin{equation} \label{eq:integratedCoherenceAfterPulse}
    \tilde{\rho}_{01}^{(t_0,\infty)} \approx \underbrace{-\frac{\frac{1}{\text{Re}\,\lambda_+}}{\left|\frac{1}{T_-} - \frac{1}{T_+}\right|}}_{=: F}|\rho_{GX}(t_0)|
\end{equation}
for the integral over the relaxation period. In the main text, Equation~(\ref{eq:integratedCoherenceAfterPulse}) was reproduced and discussed as Equation~(\ref{eq:PNCApproxFRhoGX}). The full integral is thus approximated by
\begin{equation} \label{eq:integratedCoherenceOverall}
    \tilde{\rho}_{01} \approx F|\rho_{GX}(t_0)| + \tilde{\rho}_{01}^{(-\infty,t_0)}.
\end{equation}
This result can now be analyzed by investigating the three components $F$, $|\rho_{GX}(t_0)|$ and $\tilde{\rho}_{01}^{(-\infty,t_0)}$. The latter two are obtained from the full simulation because they are relatively complicated to calculate since they depend on the laser envelope $f$. The factor $F$, however, describes the impact of the relaxation after the excitation and is therefore independent of the
details of the pulse and expressed by the simple analytical formulas given above. It is displayed in Figure~\ref{fig:finalPhotonDMs}~b) and \ref{fig:3DDifferentTempsPublication}~a).

The other components are displayed as a function of $\Delta\omega_{CX}$ in \textbf{Figure~\ref{fig:analyticFactor}}. Now the existence of the central minimum, when phonons are active, and the lack of one without phonons (red dashed line), can be explained. The factor $F$ has a minimum at \mbox{$\hbar\Delta\omega_{CX} = 0~\text{meV}$} and two maxima on either side. Because of Equation~(\ref{eq:integratedCoherenceAfterPulse}) this minimum is inherited by $\tilde{\rho}_{01}^{(t_0,\infty)}$, with and without phonons, as can be seen in Figure~\ref{fig:analyticFactor} b). However, because the initial value $|\rho_{GX}(t_0)|$ has a higher peak around \mbox{$\hbar\Delta\omega_{CX} = 0~\text{meV}$} and is overall smaller for the case without phonons, in that case, the minimum is less pronounced.

Interestingly, the integrated coherence during the pulse $\tilde{\rho}_{01}^{(-\infty,t_0)}$ appears to largely be independent of the presence of phonons. Due to the large initial coherence $|\rho_{GX}(t_0)|$, the integrated coherence during the relaxation is about one order of magnitude larger when phonons are involved, than $\tilde{\rho}_{01}^{(-\infty,t_0)}$. Because of this, the minimum can also be found in $\tilde{\rho}_{01}$. When phonons are not active, the already small minimum of $\tilde{\rho}_{01}^{(t_0,\infty)}$ essentially disappears when adding $\tilde{\rho}_{01}^{(-\infty,t_0)}$ to it, such that the central minimum of $\tilde{\rho}_{01}$ is almost imperceptible when phonons are not present.

The full simulation results are also displayed in Figure~\ref{fig:analyticFactor}. Without phonons the simple analytical model describes $\tilde{\rho}_{01}$ very well, however, with phonons, it applies to a good extend only to positive detunings. For negative $\Delta\omega_{CX}$ a large discrepancy is found that can be explained by phonon-induced decoherence. Also, the maxima appear closer to \mbox{$\hbar\Delta\omega_{CX} = 0~\text{meV}$} and the minimum is higher for the full simulation. This is likely an effect of the renormalized QD-cavity coupling $g$, because $F$ displays exactly these traits for smaller $g$.

How the factor $F$ behaves for decreasing QD-cavity coupling is displayed in Figure~\ref{fig:3DDifferentTempsPublication}~a). For \mbox{$\hbar g\approx 0.01~\text{meV}$} the two maxima at either side of the minimum at \mbox{$\hbar\Delta\omega_{CX} = 0~\text{meV}$} merge into a single maximum. As a result, the disappearing central minimum for increasing temperatures in Figure~\ref{fig:3DDifferentTempsPublication}~b) and c) from Sec.~\ref{sec:finalDMDescription} can be explained by the effective decrease in QD-cavity coupling.

\section*{Appendix B: PNC long-time decay rates}
\begin{figure}
    \centering
    \includegraphics[scale = 1]{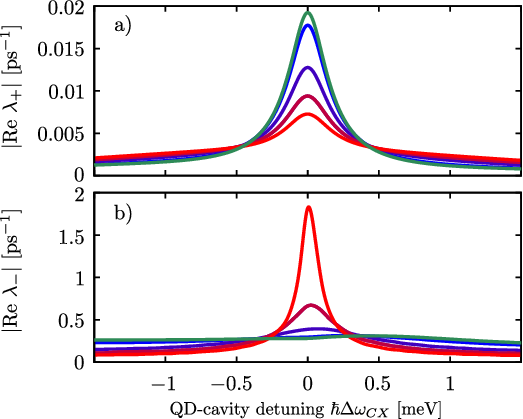}
    \caption{Absolute values of the real components of the exponential decay rates $\lambda_\pm$ of the photonic coherence $\rho_{01}$ as a function of QD-cavity detuning $\hbar\Delta\omega_{CX}$ for temperatures \mbox{4.2 K} and \mbox{10 K} to \mbox{70 K} in \mbox{20 K} steps displayed in colors of the blue-red spectrum obtained by fitting Equation~(\ref{eq:analyticApprox}) to the numerical results of the real time behavior. The case without phonons is displayed in green. a) Displays the long-time decay rate and b) shows the short-term decay rate that governs the initial drop-off.}
  \label{fig:coherenceDecayRatesPublication}
\end{figure}

The absolute  values of the real parts  of the decay rates from Equation~(\ref{eq:analyticApprox}) $\lambda_\pm$ are displayed in their dependence of the QD-cavity detuning in \textbf{Figure~\ref{fig:coherenceDecayRatesPublication}}. Both are generally negative and therefore decay the photonic coherence. These values were obtained by fitting the model developed in Appendix A to the simulated data, as it was also done in Figure~\ref{fig:fitExamples}.\\

The long-time decay rate $\lambda_+$ in subfigure~a) shows a central maximum for all temperatures (blue-red spectrum) and without phonons (green). Naturally this is due to the more efficient coupling between QD and cavity at these small detunings. What may be surprising, however, is that the height of this maximum decreases with increasing temperature. We accredit this to the phonon-induced renormalization of the QD-cavity coupling,\cite{Kaer2010,Glaessel2012,Hopfmann2015,MildeKnorrHughes2008} which was also considered in Section~\ref{sec:finalDMDescription}. In fact, similar results to the high temperature decay rates can be obtained, when plugging a smaller QD-cavity coupling $g$ into Equation~\ref{eq:ReLambda1} (not shown). An unintuitive result of this is the fact, that the long-time decay rate of the coherence decreases despite the increased effects of phonons. The general shape of a single maximum near zero-detuning is kept at all temperatures, regardless of the presence of phonons, but when phonons are active at low temperatures the curve becomes asymmetric due to phonon emission processes assisting the creation of a cavity photon in case of negative $\Delta\omega_{CX}$. For higher temperatures, the opposite process, where a phonon is absorbed, can take place as well,\cite{Glaessel2011} such that the curve becomes more symmetric with rising temperatures.\\

Figure~\ref{fig:coherenceDecayRatesPublication}~b) shows the short-time decay rates $\lambda_-$. In the phonon-free case, which is again displayed in green, only a small dip at zero detuning can be seen.  Apart from this, the rate is independent from the detuning. The addition of phonons produces not just the asymmetry also observed in subfigure~a), but interestingly also changes the general shape. For the highest displayed temperature of 70~K this is particularly obvious. The central minimum instead turns into a very pronounced maximum, which decreases to zero for large detunings. This change in shape does not change the earlier observation that $\Lambda_+$ is the more important rate when calculating $\tilde{\rho}^{(t_0,\infty)}_{GX}$, since even at 70~K and even at a QD-cavity detuning of 1.5~meV, there is still about one order of magnitude difference between the two rates. Thus, the short-time decay is not significant in our simple model of the integration when calculating $\tilde{\rho}^{(t_0,\infty)}_{GX}$.

\section*{Appendix C: PNC Rabi Rotations}
\begin{figure}
        \centering
        \includegraphics[scale = 1]{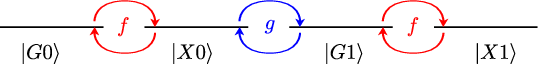}
        \caption{Energy diagram of the simplest model reproducing the $\pi/2$-alternations in the Rabi rotations of PNC.}
        \label{fig:psmModel}
    \end{figure}

When calculating Rabi rotations, a puzzling detail was found in Sec.~\ref{sec:rabiRotations}, where we observed that every other peak of PNC in its pulse-area dependence is slightly higher than the next peak.  This Appendix takes a look at a simple model, which reproduces this finding. Because the effect occurs in the phonon-free case as well, phonons can be excluded from this model. Additionally, all Lindblad-rates are discarded, because their effects are not needed to understand the observed behavior. As a result, the QD-cavity system can be described by a pure state. Under the assumption that the pulse is short enough to obtain a negligible two-photon component, all states higher than an occupation of $n=1$ may be discarded as well. To simplify the calculation, $\hbar\Delta\omega_{CX} = \hbar\Delta\omega_{LX} = 0~\text{meV}$ are used. The resulting stripped down model is schematically displayed in \mbox{Figure \ref{fig:psmModel}}. For simplicity, the laser is described as a rectangular pulse starting at $t=0~\text{ps}$ and ending at some $t_0$. Thus, the coupling between the QD's ground and excited states is given by
    \begin{equation}
        f = \frac{\Theta}{t_0}.
    \end{equation}

    Let the state be given by
    \begin{equation}
        \vert \Psi(t)\rangle = c_0(t)\vert G0\rangle + c_1(t)\vert X0\rangle + c_2(t)\vert G1\rangle + c_3(t)\vert X1\rangle,
    \end{equation}
    then the Schrödinger equation for the coordinates $c_i$, results in
    \begin{equation}
        \frac{\text{d}}{\text{d}t}\left(\begin{array}{c}c_0 \\c_1\\c_2\\c_3\end{array}\right) = -\text{i}\left(\begin{array}{cccc}
            0 & -\frac{f}{2} & 0 & 0 \\
            -\frac{f}{2} & 0 & g & 0\\
            0 & g & 0 & -\frac{f}{2}\\
            0 & 0 & -\frac{f}{2} & 0
        \end{array}\right)\cdot \left(\begin{array}{c}c_0 \\c_1\\c_2\\c_3\end{array}\right).
    \end{equation}
    Since it will be useful for later Taylor expansions, the Hamiltonian matrix is expanded with $\frac{2f}{2f}$, which results in
    \begin{equation}
        \frac{\text{d}}{\text{d}t}\left(\begin{array}{c}c_0 \\c_1\\c_2\\c_3\end{array}\right) = \text{i}\,\frac{f}{2}\left(\begin{array}{cccc}
            0 & 1 & 0 & 0 \\
            1 & 0 & -2r & 0\\
            0 & -2r & 0 & 1\\
            0 & 0 & 1 & 0
        \end{array}\right)\cdot \left(\begin{array}{c}c_0 \\c_1\\c_2\\c_3\end{array}\right),
    \end{equation}
    where $r \coloneqq g/f$ has been newly defined. Since the laser $f$ is much larger than the QD-cavity coupling $g$, this represents a system of two Rabi-oscillators, which are weakly coupled. If one solves it and varies the time $t_0$, a beating effect would be observed. But here, we want to investigate the pulse-area dependence. This problem can be solved analytically, which gives the diagonal matrix of eigenvalues $S$ and the transformation matrix into the eigenbasis $U$ to be
    \begin{align}
        S &= -\text{i}\,\frac{f}{2}\,\text{diag}(d_0, d_1, d_2, d_3)\\
        U &= \left(\begin{array}{cccc}
            A/d_0 & B/d_1 & -C/d_2 & -D/d_3 \\
            A & B & -C & -D \\
            A & B & C & D \\
            A(2r + d_0) & B(2r + d_1) & C(2r + d_2) & D(2r + d_3)
        \end{array}\right)
    \end{align}
    \begin{align}
        d_0 &\coloneqq -r + \sqrt{r^2 + 1} && d_1 \coloneqq -r - \sqrt{r^2 + 1}\\
        d_2 &\coloneqq r + \sqrt{r^2 + 1} && d_3 \coloneqq r - \sqrt{r^2 + 1}
    \end{align}
    where $A$, $B$, $C$ and $D$ are normalizing constants of the eigenvectors. Since $U$ is real and its column-vectors are normalized, its adjunct matrix is given by its transpose matrix, $U^\dagger = U^T$. The integration of the ordinary differential equation should take place for the duration of the pulse $t_0$, such that the resulting state after the excitation is given by
    \begin{equation} \label{eq:formalPSMsolution}
        \textbf{c} = U\,\text{e}^{S\,t_0}U^\dagger\textbf{c}(0),
    \end{equation}
    where the initial state is the ground state $|\Psi(0)\rangle = \vert G0\rangle$, such that $\textbf{c}(0) = (1,0,0,0)^T$. Since $d_0 = -d_3$ and $d_1 = -d_2$, as well as $A^2 = D^2$ and $B^2 = C^2$ apply, some of the exponential functions in \mbox{Equation (\ref{eq:formalPSMsolution})} merge into sines and cosines. In the arguments of those, $\Theta = ft_0$ is used, resulting in the exact solution

    \begin{equation} \label{eq:PSMExatSolution}
            \begin{split}
            \textbf{c}(\Theta) &= 2A^2\left(\begin{array}{c}
                \frac{1}{d_0^2}\,\text{cos}\left(\frac{d_0}{2}\Theta\right)\\
                \\
                \frac{\text{i}}{d_0}\,\text{sin}\left(\frac{d_0}{2}\Theta\right)\\
                \\
                \frac{1}{d_0}\,\text{cos}\left(\frac{d_0}{2}\Theta\right)\\
                \\
                \text{i}\,\left(\frac{2r}{d_0}+1\right)\,\text{sin}\left(\frac{d_0}{2}\Theta\right)
                \end{array}\right)\\
                &+ 2B^2\left(\begin{array}{c}
                \frac{1}{d_1^2}\,\text{cos}\left(\frac{d_1}{2}\Theta\right)  \\
                \\
                \frac{\text{i}}{d_1}\,\text{sin}\left(\frac{d_1}{2}\Theta\right)\\
                \\
               \frac{1}{d_1}\,\text{cos}\left(\frac{d_1}{2}\Theta\right)\\
                \\
                \text{i}\,\left(\frac{2r}{d_1}+1\right)\,\text{sin}\left(\frac{d_1}{2}\Theta\right)
            \end{array}\right)\\
        \end{split}
    \end{equation}

   Clearly, it is not very intuitive to read anything out of \mbox{Equation (\ref{eq:PSMExatSolution})}. Instead, the fact that $r \ll 1$ can be used for some approximations
    \begin{align}
        2\frac{A^2}{d_0^2} &\approx 2\frac{B^2}{d_1^2} \approx \frac{1}{2}\\
        2\frac{A^2}{d_0} &\approx \frac{1}{2}(1-r)\\
        2\frac{B^2}{d_1} &\approx -\frac{1}{2}(1+r)\\
        2A^2\left(\frac{2r}{d_0}+1\right) &\approx \frac{1}{2}(2r + 1)\\
        2B^2\left(\frac{2r}{d_1}+1\right) &\approx \frac{1}{2}(-2r + 1).
    \end{align}
    Similarly, the exponents can be expanded to give
    \begin{align}
        \frac{d_0}{2}\Theta \approx \frac{\Theta-gt_0}{2} && \frac{d_1}{2}\Theta \approx -\frac{\Theta+gt_0}{2}.
    \end{align}
    When these approximations are plugged into \mbox{Equation (\ref{eq:PSMExatSolution})}, one obtains
    \begin{equation} \label{eq:PSMApproximateSolution}
        \begin{split}
            \textbf{c}(\Theta) &= \frac{1}{2}\left(\begin{array}{c}
                \text{cos}\left(\frac{\Theta-gt_0}{2}\right) + \text{cos}\left(\frac{\Theta+gt_0}{2}\right)\\
                \\
                \text{i}\,\left[\text{sin}\left(\frac{\Theta-gt_0}{2}\right) + \text{sin}\left(\frac{\Theta+gt_0}{2}\right)\right]\\
                \\
                \text{cos}\left(\frac{\Theta-gt_0}{2}\right) - \text{cos}\left(\frac{\Theta+gt_0}{2}\right)\\
                \\
                \text{i}\,\left[\text{sin}\left(\frac{\Theta-gt_0}{2}\right) - \text{sin}\left(\frac{\Theta+gt_0}{2}\right)\right]
            \end{array}\right) \\
            &+r \left(\begin{array}{c}
                0\\
                \\
                \text{i}\,\left[-\text{sin}\left(\frac{\Theta-gt_0}{2}\right) + \text{sin}\left(\frac{\Theta+gt_0}{2}\right)\right]\\
                \\
                -\text{cos}\left(\frac{\Theta-gt_0}{2}\right) - \text{cos}\left(\frac{\Theta+gt_0}{2}\right)\\
                \\
                2\text{i}\,\left[\text{sin}\left(\frac{\Theta-gt_0}{2}\right) + \text{sin}\left(\frac{\Theta+gt_0}{2}\right)\right]
            \end{array}\right).
        \end{split}
    \end{equation}
    It is easy to see that this object is identical to the typical Rabi rotations of a two level system for $g=0$, where only the top two elements would oscillate -- one with a cosine and one with sine -- and the two states corresponding to the excited cavity would stay empty. In the following, this expression is simplified, but no further approximations are made. Using the identity
    \begin{equation} \label{eq:sinussinusIdentitaet}
        \begin{split}
            a\,\text{sin}(x+\alpha) + b\,\text{sin}(x+\beta) =\\
            \sqrt{a^2+b^2+2ab\,\text{cos}(\alpha-\beta)}\,\text{sin}(x+\delta),
        \end{split}
    \end{equation}
    with
    \begin{equation}
        \delta = \text{atan2}(a\,\text{sin}\,\alpha + b\,\text{sin}\,\beta, a\,\text{cos}\,\alpha + b\,\text{cos}\,\beta),
    \end{equation}
    and using
    \begin{align}
        \Tilde{A}\coloneqq\left\vert\text{cos}\left(\frac{gt_0}{2}\right)\right\vert &= \frac{\sqrt{2+2\,\text{cos}(gt_0)}}{2}\\
        \Tilde{B}\coloneqq\left\vert\text{sin}\left(\frac{gt_0}{2}\right)\right\vert &= \frac{\sqrt{2-2\,\text{cos}(gt_0)}}{2},
    \end{align}
    the linear combinations in \mbox{Equation (\ref{eq:PSMApproximateSolution})} can be rewritten as
    \begin{equation}
         \textbf{c}(\Theta) = \frac{1}{2}\left(\begin{array}{c}
           \Tilde{A}\,\text{cos}\left(\frac{\Theta}{2}\right)\\
            \\
            \text{i}\Tilde{A}\,\text{sin}\left(\frac{\Theta}{2}\right)\\
            \\
            \Tilde{B}\,\text{sin}\left(\frac{\Theta}{2}\right)\\
            \\
            -\text{i}\Tilde{B}\,\text{cos}\left(\frac{\Theta}{2}\right)
        \end{array}\right) + r \left(\begin{array}{c}
            0\\
            \\
            \text{i}\Tilde{B}\,\text{cos}\left(\frac{\Theta}{2}\right)\\
            \\
            -\Tilde{A}\,\text{cos}\left(\frac{\Theta}{2}\right)\\
            \\
            2\text{i}\Tilde{A}\,\text{sin}\left(\frac{\Theta}{2}\right)
        \end{array}\right).
    \end{equation}
    Similar to this step, another identity
    \begin{equation}
        \begin{split}
            a\,\text{cos}\,\alpha + b\,\text{sin}\,\alpha =\\ \text{sgn}(a)\sqrt{a^2+b^2}\,\text{cos}\left(\alpha + \text{atan}\left(-\frac{b}{a}\right)\right),
        \end{split}
    \end{equation}
    can be used to combine these sines and cosines further and avoid writing the $c_j$ as linear combinations. The resulting amplitudes $X,Y,Z$ and phase angles $\delta_1, \delta_2$ and $\delta_3$ are only weakly dependent on changes in $\Theta$ within any $[n\pi,(n+2)\pi]$ interval and may thus, for the purposes of this investigations be considered constant, since the main point of this Appendix is to investigate the PNC's behavior within such an interval. The result is
    \begin{equation}
         \textbf{c}(\Theta) = \frac{1}{2}\left(\begin{array}{c}
           \Tilde{A}\,\text{cos}\left(\frac{\Theta}{2}\right)\\
            \\
            \text{i}X(\Theta)\,\text{sin}\left(\frac{\Theta}{2} + \delta_1(\Theta)\right)\\
            \\
            -Y(\Theta)\,\text{sin}\left(\frac{\Theta}{2} + \delta_2(\Theta)\right)\\
            \\
            -\text{i}Z(\Theta)\,\text{cos}\left(\frac{\Theta}{2} + \delta_3(\Theta)\right)
        \end{array}\right).
    \end{equation}
    
    Now the PNC just after the pulse can be written as
    \begin{equation} \label{eq:psmPNC}
        \begin{split}
            \rho_{01}(\Theta) &= c_0^*c_2 + c_1^*c_3\\
            &= -\Tilde{A}Y\underbrace{\text{cos}\left(\frac{\Theta}{2}\right)\text{sin}\left(\frac{\Theta}{2}+\delta_2\right)}_{=[\text{sin}(\delta_2)\,+\,\text{sin}(\Theta+\delta_2)]/2}\\
            &- XZ\underbrace{\text{cos}\left(\frac{\Theta}{2}+ \delta_3\right)\text{sin}\left(\frac{\Theta}{2}+\delta_1\right)}_{=[\text{sin}(\delta_1-\delta_3)\,+\,\text{sin}(\Theta+\delta_1+\delta_3)]/2}\\
            &= -\underbrace{\frac{\Tilde{A}Y}{2}\,\text{sin}\left(\Theta+\delta_2\right) - \frac{XZ}{2}\,\text{sin}\left(\Theta+\delta_1+\delta_3\right)}_{\stackrel{(\ref{eq:sinussinusIdentitaet})}{=:}W\,\text{sin}(\Theta+\sigma)}\\
            &- \underbrace{\frac{\Tilde{A}Y}{2}\,\text{sin}\left(\delta_2\right) - \frac{XZ}{2}\,\text{sin}\left(\delta_1-\delta_3\right)}_{=: O},
        \end{split}
    \end{equation}
    where $\sigma$ and
    \begin{equation}
        W \coloneqq \frac{1}{2}\sqrt{\Tilde{A}^2Y^2+X^2Z^2+2\Tilde{A}XYZ\,\text{cos}(\delta_1-\delta_2+\delta_3)}    
    \end{equation}

    \begin{figure}
        \centering
        \includegraphics[scale = 1]{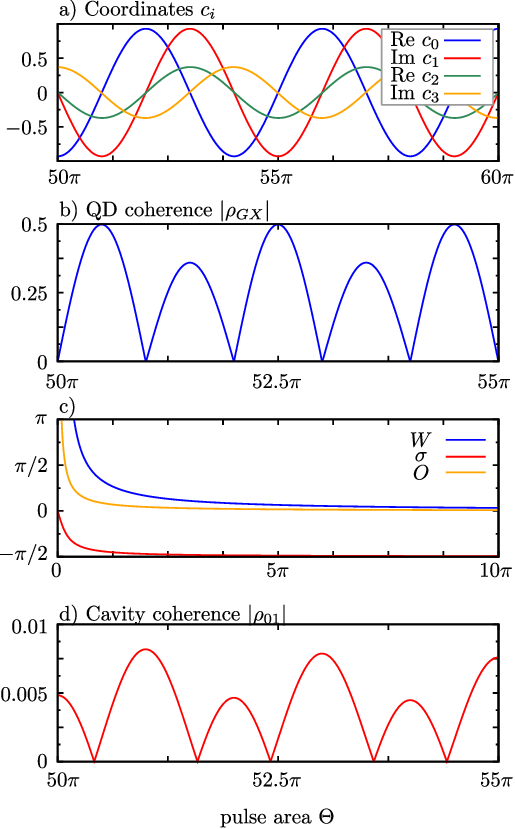}
        \caption{Results of Appendix C's calculations. The QD-cavity coupling was chosen as $g = 0.05~\text{meV}$ and the pulse-length is $t_0=10~\text{ps}$.}
        \label{fig:psmFactors}
    \end{figure}
    
    result out of \mbox{Equation (\ref{eq:sinussinusIdentitaet})} again. Thus, the coherence can be written as a sine, which has a small phase angle and an offset. As one would expect, the frequency with which the coherence oscillates is double that of the population. When taking the absolute value, the offset produces the difference between every other PNC peak. These results are plotted in \mbox{Figure \ref{fig:psmFactors}}, which displays the coordinates $c_i$, the absolute QD coherence, which was calculated calculated via $|\rho_{GX}| = |c_0^*c_1 + c_2^*c_3|$, the parameters $W$, $\sigma$ and $O$ from \mbox{Equation (\ref{eq:psmPNC})} and PNC, which was calculated using the absolute value of
    \begin{equation}
        \rho_{01} = W\,\text{sin}(\Theta + \sigma) + O.
    \end{equation}
    The coordinates $c_0$ and $c_1$ behave very similarly to the usual driven two-level model. The states with an excited cavity are -- with the chosen parameters $gt_0=0.7596$ -- much less occupied. They oscillate with the same frequency, but they posses shifted phases. These phases converge $\delta_2 \to \pi$ and $\delta_3 \to -\pi/2$ for large $\Theta$. The last phase very quickly converges against $\delta_1 \to 0$. These phase differences, especially $\delta_3$, which converges the slowest, create the offset $O$ in \mbox{Equation (\ref{eq:psmPNC})} and thus result in the observed phenomenon. \mbox{Figure \ref{fig:psmFactors} c)} may suggest that the offset decreases, out of which one might conclude that the effect decreases. It does not, since the amplitude $W$ decreases roughly as quickly. This can be seen in the fact that \mbox{Figure \ref{fig:psmFactors} b)} and d) display the coherences for $\Theta\in[50\pi,55\pi]$, where $O$ and $W$ have largely converged.
    
\bibliographystyle{MSP}
\bibliography{bibfile.bib}

\end{document}